\documentclass{aa}  

\usepackage{graphicx}
\usepackage{txfonts}

\begin{document}

   \title{Post-Newtonian gravity and Gaia-like astrometry}

   \subtitle{Effect of PPN $\gamma$ uncertainty on parallaxes}

   \author{Alexey G. Butkevich
          \inst{1,2}
          \and
          Alberto Vecchiato
          \inst{1}
          \and
          Beatrice Bucciarelli
          \inst{1}
          \and
          Mario Gai
          \inst{1}
          \and
          Mariateresa Crosta
          \inst{1}
          \and
          Mario~G.~Lattanzi
          \inst{1}
          }

   \institute{INAF - Astrophysical Observatory of Torino, 10025 Pino Torinese, Italy\\
              \email{alexey.butkevich@inaf.it}
         \and
             Pulkovo Observatory, Russian Academy of Sciences, 196140 Saint Petersburg, Russia}

   \date{Received ...; accepted ...}

 
  \abstract
   {Relativistic models of light propagation adopted for high-precision astrometry are based on the parametrised post-Newtonian formalism, which provides a framework for examining the effects of a hypothetical violation of general relativity on astrometric data. Astrometric observations are strongly affected by the post-Newtonian parameter $\gamma$ describing the strength of gravitational light deflection.}
   {We study both analytically and numerically how a deviation in the PPN parameter $\gamma$ from unity, which is the value predicted by general relativity, affects the parallax estimations in \emph{Gaia}-like astrometry.}
   {Changes in the observable quantities produced by a small variation in PPN $\gamma$ were calculated analytically. We then considered how such variations of the observables are reflected in the parallax estimations, and we performed numerical simulations to check the theoretical predictions.}
   {A variation in the PPN $\gamma$ results in a global shift of parallaxes and we present a formula describing the parallax bias in terms of the satellite barycentric distance, the angle between the spin axis and the direction to the Sun, and the PPN $\gamma$ uncertainty. Numerical simulations of the astrometric solutions confirm the theoretical result. The up-to-date estimation of PPN $\gamma$ suggests that a corresponding contribution to the \emph{Gaia} parallax zero point unlikely exceeds 0.2 $\mu$as. The numerical simulations indicate that the parallax shift is strongly dependent on ecliptic latitude. It is argued that this effect is due to an asymmetry in the \emph{Gaia} scanning law and this conclusion is fully validated by additional simulations with a reversed direction of the precession of the spin axis around the direction to the Sun.}
   {}

   \keywords{gravitation -- methods: data analysis -- methods: statistical -- space vehicles: instruments -- astrometry}

   \maketitle
%

\section{Introduction}
Among the goals of the ESA \emph{Gaia} mission \citep{2016A&A...595A...1G}, probably the most fundamental one is to produce an absolute astrometric catalogue, which is a realisation of a quasi-inertial reference system on the sky, down to magnitude $G=20.7$, and with accuracies ranging from 10 to 1000 microarcseconds ($\mu$as) for more than one billion stars.
A real breakthrough for astrometry, with a profound impact on many fields of astronomy and astrophysics, is the availability of absolute parallaxes at $\sim10\%$ accuracy on a Galaxy-wide scale. The estimation of these quantities is the main goal of absolute astrometry, in contrast to relative astrometry, which is usually able to determine parallaxes only with respect to background objects, and thus it is affected by  systematic errors that depend on these background references.

Reaching this goal with \emph{Gaia}, however, is potentially hampered by the coupling between the parallax and other parameters, which can result in a bias between the computed and true parallaxes. This systematic error is generally refered to as the parallax zero point.

Some of these correlations are intrinsically tied to the way \emph{Gaia} takes its measurements; others have a more fundamental character and can affect astrometric measurements in more general cases.
The coupling of a global parallax shift with variations in the basic angle between the two fields of view (FOVs) \citep{Butkevich+2017} as well as the correlation between the parallax and the rate of  across-scan motion of sources in the FOVs \citep{2021A&A...649A...2L} illustrate the correlations of the first type.

The second type of coupling is exemplified well by the correlation between parallaxes and the so-called parameter $\gamma$, which is one of the parameters utilised in the parametrised post-Newtonian (PPN) formalism \citep{2014LRR....17....4W,Poisson+Will2014}. This framework allows one to classify and quantify deviations in a gravity theory from general relativity (GR) at its first post-Newtonian order. The signature of such deviations is given by the values of a set of ten parameters, each associated to a specific kind of difference and related to one relativistic effect or more.

The $\gamma$ parameter of the PPN formalism is related to the amount of space curvature generated by a unit of mass-energy, and for this reason it directly enters in the formulae that describe the gravitational light bending. For this reason, astrometric measurements are well suited to test GR through the estimation of this PPN parameter.

This task was pursued in the past \citep{1989racm.book.....S}, but its effectiveness could only significantly improve with the advent of space astrometry, thanks to a huge increase in the measurement accuracy, number of observed objects, and observing time. Space-based astrometric missions have been proposed many times in the past, but currently only two of them have actually been realised: HIPPARCOS \citep{1997ESASP1200.....E} and \emph{Gaia}.

They are both survey missions based on the same sky-scanning principle, and for which the most direct way to estimate the value of $\gamma$ is to include it among the unknowns of the so-called global astrometric sphere reconstruction \citep{lindegrenAGIS2012,Vecchiato+2018}. In this way, the parameter comes out from the reconstruction of the celestial global reference system together with the standard astrometric parameters, namely including the positions, parallaxes, and proper motions.

Therefore, the task of the sphere reconstruction turns into a huge test of gravity theories at the same time, and this is exactly what is planned for \emph{Gaia}. On the other hand, its forerunner, HIPPARCOS, adopted a strategy based on an a posteriori estimation of $\gamma$ from the residuals of the sphere reconstruction \citep{Froeschle+1997}.

From what was mentioned above, however, it is clear that the existing correlation between parallaxes and the PPN parameter $\gamma$ can jeopardise the scientific results of an astrometric mission  in many
ways, and of a \emph{Gaia}-like one in particular. For example, a systematic bias on the parallaxes could undermine the accuracy of the PPN $\gamma$ parameter coming from the sphere reconstruction. Or, alternatively, an error on the latter might result in biased parallaxes.

 Such entanglement, from the point of view of a global astrometric mission, is extremely worrisome because it can put the determination of absolute parallaxes at stake, that is the most important scientific goal of these kinds of missions. In this sense, it is worth mentioning that the estimated global parallax zero point amounts to about $-17~\mu\mathrm{as}$ in the recently published \emph{Gaia} Early Data Release 3 (EDR3) \citep{LindegrenEDR3Astr2020}.

General relativity is one of the cornerstones of modern physics and its predictions have been confirmed in numerous experiments; alternative theories of gravity are, nevertheless, thought to be viable \citep{2014LRR....17....4W}. The effects of a deviation from GR on the outcome of a \emph{Gaia}-like astrometric experiment have not been systematically studied so far.
This paper aims to investigate such issues, with particular attention being paid to the impact of a biased PPN $\gamma$ on the determination of parallaxes in a \emph{Gaia}-like mission.

The outline of the paper is as follows. Section~\ref{sec:theory} gives an analytical treatment for the problem. In Sect.~\ref{sec:simulations} we present the results of the numerical simulations that confirm the theoretical expectation. Section~\ref{sec:asymmetry} explores the dependence of the parallax bias on the star position. The results are discussed in Sect.~\ref{sec:conclusions}.

\section{\label{sec:theory}Theory}

The correlation between parallaxes and the PPN-$\gamma$ parameter arises from the similarity in the way the light deflection and parallax affect observed directions. An increasing light deflection (i.e. a small increase in PPN $\gamma$) results in a shift in the direction outwards from the centre of the deflecting body, whereas rising parallaxes lead to a direction shift towards the Solar System barycentre (SSB). The Sun is responsible for the most significant part of the light deflection. For a scanning mission, utilising the \emph{Gaia}-like design, large angular separations are kept between the Sun and objects of observations; for example, the \emph{Gaia} observations are carried out 45\degr \ to 135\degr \ away from the Sun. In such case, the great circles from the observed source to the solar centre and to the SSB are almost coincident and the effects of the light deflection and parallax are, therefore, nearly parallel and opposite in direction (Fig.~\ref{fig:alShift}).

This intuitive explanation is more thoroughly detailed in the following subsections, where
we provide a theoretical background
and derive some useful formulae. 
An analysis of the effect of the trigonometric parallax and PPN $\gamma$ on observables shows a strong correlation between these two quantities, in agreement with earlier results. 
To further test the influence of such correlation, we simulated Gaia-like observations using  a perturbed $\gamma$ value, and then we performed a  least-squares solution of the sole astrometric parameters by  assuming the GR value of $\gamma$. Finally, we used this approach to determine how a systematic parallax error biases the estimation of the parameter $\gamma$ and vice versa.
For brevity, all distances are expressed in the astronomical unit $A$ hereafter.

\subsection{Basic equations}

We first calculated variations in the direction to a star due to small changes in the  trigonometric parallax $\varpi$ and PPN parameter $\gamma$. The relationship between the direction $\vec{u}$ to a star and its astrometric parameters is described by the astrometric model. This model, given by Eq.~(4) of \cite{lindegrenAGIS2012}, suggests that if the parallax receives a small perturbation $\delta\varpi$, the direction is changed by
\begin{equation}\label{eq:du_par}
  \delta\vec{u}_\varpi=\vec{u}\times\left(\vec{u}\times\vec{R}\right)\delta\varpi\,,
\end{equation}
where $\vec{R}$ is the position of the observer with respect to the SSB.

In the framework of the PPN formalism, the light deflection by the solar gravitational field is given by (see, for example, \cite{lindegren+1992}, \cite{klioner2003} and \cite{Poisson+Will2014}, Sect.~13.3.2)
\begin{equation}\label{eq:deflection}
  \Delta\vec{u}=-\vec{u}\times\left(\vec{u}\times\vec{R}_\mathrm h\right)
  \frac{1+\gamma}{2}
  \frac{r_\mathrm{s}}{R_\mathrm h\left(R_\mathrm h+\vec{u}'\vec{R}_\mathrm h\right)}\,,
\end{equation}
where $r_\mathrm{s}\equiv 2GM_\sun/Ac^2=1.974\times 10^{-8}\,\mathrm{au}$ is the Schwarzschild radius of the Sun, $\vec{R}_\mathrm h$ is the heliocentric position of the observer, and the prime signifies vector scalar multiplication. As the Sun orbits the barycentre, the barycentric distance of the solar centre does not exceed 0.01 au \citep{Perryman+Schulze-Hartung2011}. Therefore, $\vec{R}$ and $\vec{R}_\mathrm h$ differ by, at most, 0.6\degr\ in direction and 1\% in length for a satellite operating in the vicinity of Earth or near the L2 point. Such a small difference suggests that we can neglect the distinction between the barycentric and heliocentric position. 
For this reason, we use $\vec{R}$ instead of $\vec{R}_h$ in what follows.

Thus, we found that the variation in the observed direction due to a small change in PPN $\gamma$ can be written as
\begin{equation}\label{eq:du_gam}
  \delta\vec{u}_\gamma=-\vec{u}\times\left(\vec{u}\times\vec{R}\right)\frac{r_\mathrm{s}}{2R^2}\,\frac{\delta\gamma}{1-\cos\chi}\,,
\end{equation}
with $\chi$ being the angular distance between the Sun and the star as seen by the observer, $\vec{u}'\vec{R}=-R\cos\chi$. The minus sign is present because $\vec{R}$ is the vector from the barycentre to the observer, while $\vec{u}$ points from the observer towards the star.

The barycentric distance $R$ is not strictly constant but slightly time-dependent. For instance, the \emph{Gaia} barycentric distance ranges from 0.99 to 1.03 au. We regard $R$ as a constant value assuming  $R=1.01\,\mathrm{au}$.

\subsection{Gaia-like observations}
\label{ss:observations}

\begin{figure}
 \resizebox{\hsize}{!}{\includegraphics{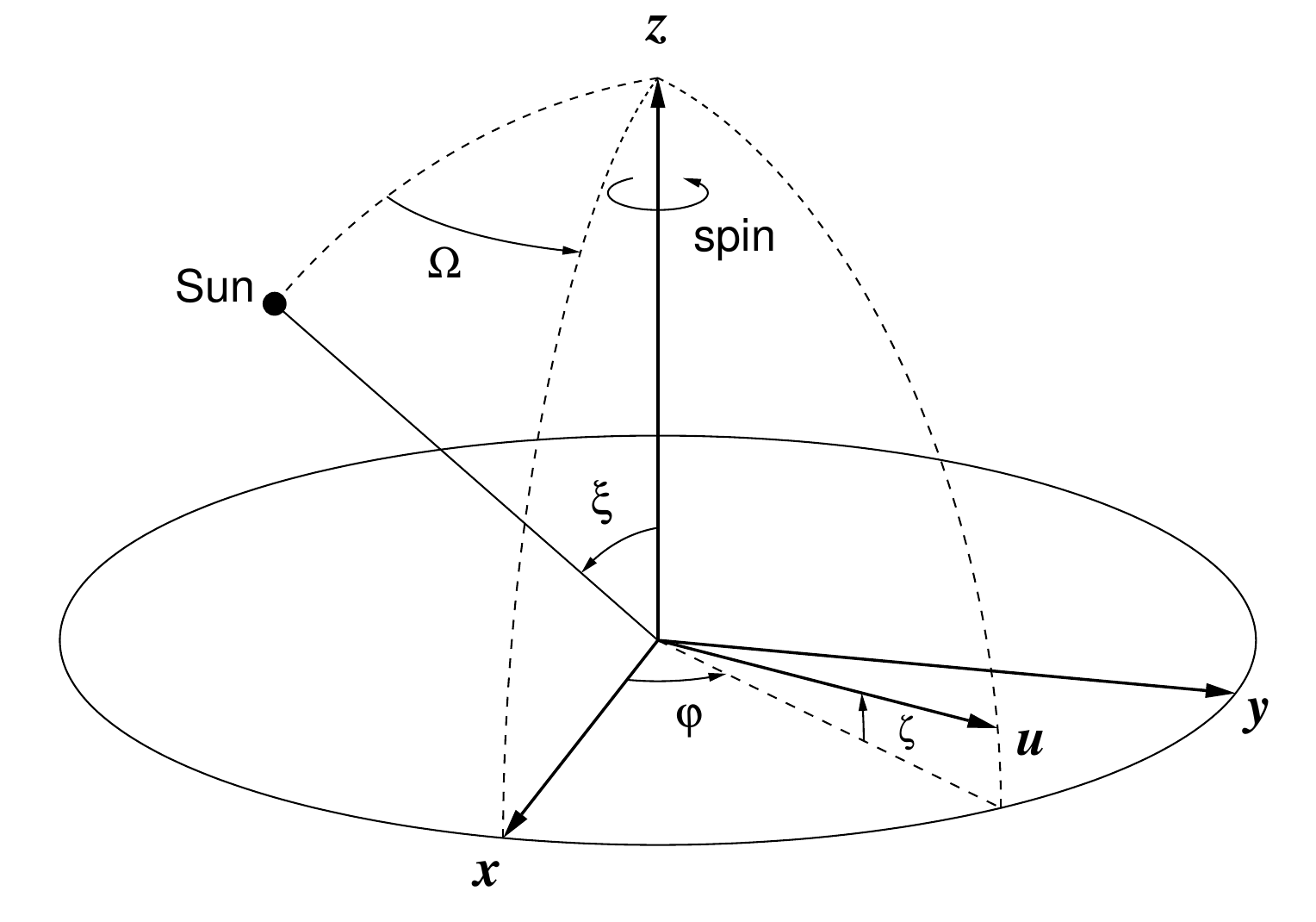}}
 \caption{Scanning reference system (SRS). The axes $\vec{x}$ and $\vec{y}$ define the instantaneous scanning circle, with the axis $\vec{z}$ being along the spin axis. The heliotropic spin phase $\Omega$ and solar aspect angle $\xi$ specify the direction towards the Sun. The observed direction $\vec{u}$ is given by the along- and across-scan instrumental angles $\varphi$ and $\zeta$, repectively.}
 \label{fig:refframe}
\end{figure}

\emph{Gaia}-like observations are conveniently described in terms of the so-called scanning reference system (SRS) \citep{lindegrenAGIS2012}. The reference system and relevant spherical coordinates are illustrated in Fig~\ref{fig:refframe}.

The SRS is defined by the instrument axes $\vec{x}$, $\vec{y}$, and $\vec{z}$. The axis $\vec{x}$ bisects the basic angle between the two viewing directions, $\vec{z}$ is the direction along the spin axis, and $\vec{y}=\vec{z}\times\vec{x}$ completes the right-handed triad. The spherical coordinates $\varphi$ and $\zeta$, referred to as the instrument angles, specify the direction towards the star:
\begin{equation}\label{eq:u}
 \vec{u}=\vec{x}\cos\varphi\cos\zeta+\vec{y}\sin\varphi\cos\zeta+\vec{z}\sin\zeta\,.
\end{equation}
Similarly, the two angles $\Omega$ and $\xi$ specify the direction to the Sun in the SRS.  The solar aspect angle $\xi$, that is the angle between the spin axis and the direction to the Sun, is almost constant. 
For \emph{Gaia}, $\xi$ was set to 45\degr; a slightly smaller value of 43\degr\ was chosen for Hipparcos.
The spin phase $\Omega$, being the angle between the direction to the Sun and the plane $xz$, increases with time at a nearly constant rate. Thus, having considered the SSB coincident with the centre of the Sun,  the barycentric position of the satellite in the SRS is
\begin{equation}\label{eq:R}
\vec{R}=R\left(-\vec{x}\cos\Omega\sin\xi+\vec{y}\sin\Omega\sin\xi-\vec{z}\cos\xi\right)\,.
\end{equation}
This equation takes into account the fact that the vector $\vec{R}$ is from the barycentre to the satellite.

The dot product of $\vec{u}$ and $\vec{R}$ gives the angular separation between the Sun and it observed the star in terms of the SRS:\begin{equation}\label{eq:cos_chi}
        \cos\chi=-\vec{u}'\vec{R}/R=\sin\zeta\cos\xi+\cos\zeta\sin\xi\cos\theta\,.
\end{equation}
Here, $\theta=\Omega+\varphi$ is the along-scan coordinate, that is the angle between the Sun and $\vec{u}$ measured around the $\vec{z}$ axis in the direction of the satellite spin. Disregarding the small across-scan extent of the FOVs, it shows that a scanning satellite observes objects $90\degr-\xi$ to $90\degr+\xi$ away from the Sun. For \emph{Gaia}, with $\xi=45\degr$, the angular separation ranges from 45\degr \ to 135\degr.

Linearising Eq.~(\ref{eq:u}) with respect to $\varphi$, we found that the change in the direction due to a small variation $\delta\varphi$ is
\begin{equation}\label{eq:du_phi}
\delta\vec{u}=\left(-\vec{x}\sin\varphi\cos\zeta+\vec{y}\cos\varphi\cos\zeta\right) \delta\varphi=\left(\vec{z}\times\vec{u}\right)\delta\varphi\,.
\end{equation}
We neglect the finite size of the FOV and simply assume that the observations are only performed on the scanning circle $\left(\zeta=0\right)$ in what follows.
For \emph{Gaia}, with the across-scan FOV size being less than 1\degr, this assumption is valid to within 1\% \citep{Butkevich+2017}.
In this approximation, the cross product $\vec{z}\times\vec{u}$ represents a unit vector along the spin direction. Taking the dot product of Eq.~(\ref{eq:du_phi}) and $\vec{z}\times\vec{u}$, we obtained the variation in the along-scan coordinate due to a change in the direction
\begin{equation}\label{eq:dphi}
\delta\varphi=\left(\vec{z}\times\vec{u}\right)'\delta\vec{u}\,.
\end{equation}

Substituting the variations of the observed direction from Eqs.~(\ref{eq:du_par}) and (\ref{eq:du_gam}) into Eq.~(\ref{eq:dphi}), we obtained the shifts of the along-scan coordinate due to changes in the parallax and PPN $\gamma$
\begin{align}
 \label{eq:dphi_par}
 \delta\varphi_\varpi&=-R\sin\xi\sin\theta\,\delta\varpi\,,\\
 \label{eq:dphi_gam}
 \delta\varphi_\gamma&=\frac{r_\mathrm s}{2R}\frac{\sin\xi\sin\theta}{1-\sin\xi\cos\theta}\,\delta\gamma\,.
\end{align}
Deriving Eq.~(\ref{eq:dphi_gam}), we used $\cos\chi$ from Eq.~(\ref{eq:cos_chi}) with $\zeta=0$.

Equations~(\ref{eq:dphi_par}) and (\ref{eq:dphi_gam}) describe the projections of displacement of the stellar image on the scanning circle. Similar formulae were derived by \cite{Froeschle+1997} and \cite{Hobbs+2009} from geometrical considerations.

\subsection{Size of the effects}

\begin{figure}
 \resizebox{\hsize}{!}{\includegraphics{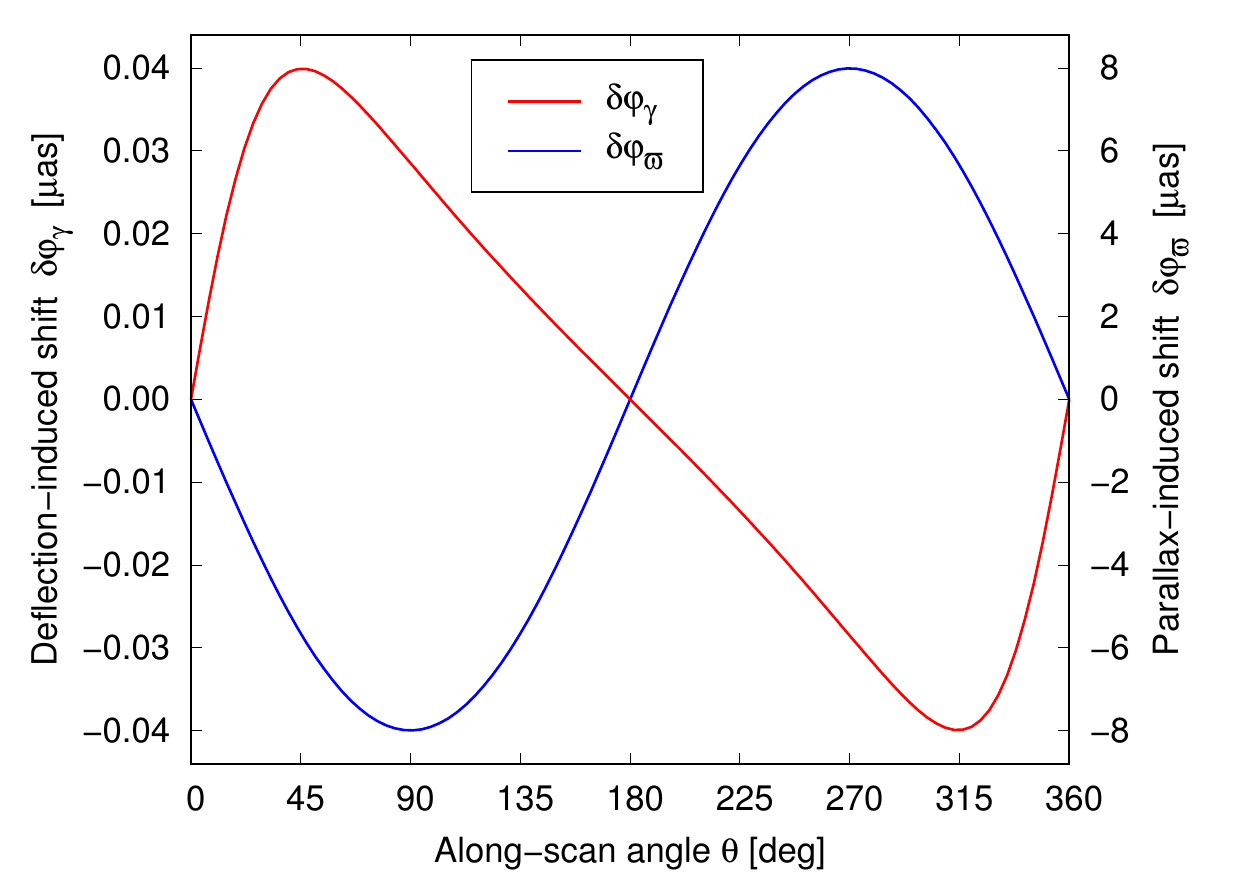}}
 \caption{Effect of changes in the light deflection and parallax on the along-scan coordinate $\varphi$. The red line, calculated from Eq.~(\ref{eq:dphi_gam}) with $\delta\gamma=2\times10^{-5}$, shows the maximum shift expected under current constraints on PPN $\gamma$. The blue line, computed using Eq.~(\ref{eq:dphi_par}), demonstrates how $\delta\varphi_\varpi$ varies with $\theta$ if the parallax increases by 11.3 $\mu$as.}
 \label{fig:alShift}
\end{figure}

The most accurate estimate of PPN $\gamma$ as of yet, which was obtained from Doppler tracking of the Cassini spececraft \citep{Bertotti+2003,2014LRR....17....4W}, is $\gamma-1=\left(2.1\pm2.3\right)\times10^{-5}$. This places a constraint on the expected variation in observables because the absolute value of $\delta\gamma$ is not likely to be much larger than $2\times10^{-5}$. The shift $\delta\varphi_\gamma$ corresponding to this value of $\delta\gamma$ is shown in Fig.~\ref{fig:alShift} as a function of the along-scan angle $\theta$, together with the shift due to parallax.

The close similarity between the effects visible in Fig.~\ref{fig:alShift} suggests a strong correlation between them. Considering $\delta\varphi_\varpi$ and $\delta\varphi_\gamma$ as functions $\theta$, \cite{Hobbs+2009} 
analytically demonstrated that the coefficient of correlation between the along-scan components of the light deflection and parallactic effects is 0.91 for the \emph{Gaia} solar aspect angle.

We now estimate the size of the effects. The maximum absolute value of the parallactic one is $R\sin\xi\,\delta\varpi$, while the maximum absolute value of the along-scan light deflection,
attained at $\theta=90\degr-\xi$ or $\theta=270\degr+\xi$, 
is $\left(r_\mathrm{s}/2R\right)\tan\xi\,\delta\gamma$. Assuming $R=1$ and keeping in mind that the effect is of an opposite sign, we see that the maximum effects are equal when
\begin{equation}
 -2\cos\xi\,\delta\varpi\simeq r_\mathrm s\,\delta\gamma\,.
\end{equation}
Expressing parallax in $\mu$as and $\gamma$ in units of $10^{-5}$, we find that, for the \emph{Gaia} solar aspect angle of 45\degr, the following relation holds 
\begin{equation}\label{eq:proportion}
 \frac
 {\delta\varpi\,\left[\mu\mathrm{as}\right]}
 {\delta\gamma\,\left[10^{-5}\right]}\simeq
 -\frac{1}{35}\,.
\end{equation}
This proportion illustrates the relative values of the effects. If, for instance, a parallax increase of 10\,$\mu$as occurs, this corresponds to a variation in the observables, which can be attributed to a $\sim350\times10^{-5}$ drop in PPN $\gamma$. Conversely, a variation in the observables resulting from a $10^{-5}$ rise in PPN $\gamma$ can be described by a parallax descrease of $\sim1/35\,\mu\mathrm{as}$, that is to say $\sim30\,\mathrm{nas}$.

This treatment is applicable to the parallax of any star, and therefore $\delta\varpi$ can be interpreted as a global shift of the parallaxes. A systematic deviation in parallaxes from their true values is referred to as a parallax zero point. Equation~(\ref{eq:proportion}) thus relates the PPN $\gamma$ uncertainty and the parallax zero point. Although this relation is approximate, it is convenient for order-of-magnitude estimations. A more accurate formula for the effect of PPN $\gamma$ on the parallax zero point is derived below in Sect.~\ref{ss:gamonpar}.

\subsection{Impact of PPN $\gamma$ on the parallax zero point}
\label{ss:gamonpar}

The variation in the along-scan instrument angle $\varphi$ due to a change in the parallax of a star is given by Eq.~(\ref{eq:dphi_par}). Considering all observations of a given star, this relation can be written in the matrix form as follows:
\begin{equation}\label{eq:p_dpi}
 \vec{p}\,\delta\varpi=\delta\vec{\varphi}\,,\end{equation}
where $\vec{p}$ and $\delta\vec{\varphi}$ are the vectors of length $n$, the number of observations of the star in question. The elements of vector $\vec{p}$ correspond to individual observations, labelled by the index $l$:
\begin{equation}
 p_l=-R\sin\xi\sin\theta_l\,.
\end{equation}

Equation~(\ref{eq:p_dpi}) represents a system of $N$ linear equations with one unknown $\delta\varpi$. To find its solution in the sense of a least-squares estimation, we first constructed the  normal equation by computing its scalar product with vector $\vec{p}$:
\begin{equation}
 \vec{p}'\vec{p}\,\delta\varpi=\vec{p}'\delta\vec{\varphi}\,.
\end{equation}
The solution of the normal equation gives a least-squares estimate of the parallax shift
\begin{equation}\label{eq:dpar_general}
 \delta\varpi=\frac{\vec{p}'\delta\vec{\varphi}}{\vec{p}'\vec{p}}\,. 
\end{equation}
This formula is generally valid because it is derived without making any assumptions concerning the variations $\delta\vec{\varphi}$. We now apply it to the case of variations due to an uncertainty in the light deflection parameter $\gamma$.

According to Eq.~(\ref{eq:dphi_gam}), variations of the along-scan instrument angle due to a change in PPN $\gamma$ are
\begin{equation}\label{eq:dphi_gam_vec}
 \delta\vec{\varphi}=\vec{g}\,\delta\gamma\,,
\end{equation}
where the entries of the vector $\vec{g}$ are
\begin{equation}
 g_l=\frac{r_\mathrm{s}}{2R}\frac{\sin\xi\sin\theta_l}{1-\sin\xi\cos\theta_l}\,.
\end{equation}
Substituting Eq.~(\ref{eq:dphi_gam_vec}) into Eq.~(\ref{eq:dpar_general}), we find an estimation of the parallax bias
\begin{equation}
 \delta\varpi_\gamma=\frac{\vec{p}'\vec{g}}{\vec{p}'\vec{p}}\,\delta\gamma\,.
\end{equation}
It is convenient to use relevant mean values instead of the scalar products $\vec{p}'\vec{p}$ and $\vec{p}'\vec{g}$:
\begin{equation}
 \vec{p}'\vec{p}=\sum_{l=1}^np_l^2=n\left\langle p^2\right\rangle\,,\quad \vec{p}'\vec{g}=\sum_{l=1}^np_lg_l=n\left\langle pg\right\rangle\,.
\end{equation}
Replacing the averaging over observations with averaging over the along-scan angle, $\left\langle x\right\rangle=\left(1/2\pi\right)\int_0^{2\pi}x\left(\theta\right)\,\mathrm d\theta$, we find 
\begin{align}
 \left\langle p^2\right\rangle&=R^2\frac{\sin^2\xi}{2}\,,\\
 \left\langle pg\right\rangle&=-r_\mathrm{s}\sin^2\left(\xi/2\right)\,.
\end{align}
Finally, we find the parallax shift
\begin{equation}\label{eq:parShift}
\begin{aligned}
        \delta\varpi_\gamma&=-\frac{r_\mathrm s}{2R^2}\frac{\delta\gamma}{\cos^2\left(\xi/2\right)}=-\frac{GM}{Ac^2}\left(\frac{A}{R}\right)^2\frac{\delta\gamma}{\cos^2\left(\xi/2\right)}\\
        &=-0.023\,\mu\mathrm{as}\times\left(\frac{\delta\gamma}{10^{-5}}\right)\quad\mbox{for }R=1.01\mbox{ and }\xi=45\degr\,.
\end{aligned}
\end{equation}
This formula gives the solution to the problem. It determines the parallax zero point originating from a PPN $\gamma$ uncertainty for \emph{Gaia}-like astrometry. According to today's best estimate by \citet{Bertotti+2003}, a 3-$\sigma$ deviation of $\gamma-1$ from the GR value of zero amounts to $9\times10^{-5}$. The expected effect of the parallax zero point is 0.2\,$\mu$as. This tiny effect is truly negligible for the \emph{Gaia} astrometry.

Equation~(\ref{eq:parShift}) presents the sky-averaged parallax bias. The bias is furthermore subject to zonal variations due to different observational configurations for various star positions. This is well exemplified by areas close to the ecliptic poles. Straightforward geometrical arguments suggest that such stars are observed only when $\theta\simeq 90\degr$ or 270\degr. Accordingly, in this case, $\vec{p}'\vec{p}\simeq R^2\sin^2\xi$ and $\vec{p}'\vec{g}\simeq-\left(r_\mathrm{s}/2\right)\sin^2\xi$ and the parallax shift becomes 
\begin{equation}
        \delta\varpi_\gamma\simeq-\frac{r_\mathrm s}{2R^2}\,\delta\gamma
        =-0.020\,\mu\mathrm{as}\times\left(\frac{\delta\gamma}{10^{-5}}\right)\quad\mbox{for }R=1.01\,.
\end{equation}
Thus the parallax bias is independent of the solar aspect angle near the ecliptic poles. We further address the bias dependence of star position in Sect.~\ref{sec:asymmetry}.

\section{Monte Carlo experiments}
\label{sec:simulations}

To test the consistency of the theory, numerical tests were conducted. In the following, we first describe the basic computational tasks, that is the  generation of simulated \emph{Gaia}-like observations and the astrometric solution, which uses a numerical technique to estimate astrometric parameters from the simulated observations. We then employ a statistical approach to assess the main characteristics of the simulations outcome. Finally, we present the results of the Monte Carlo experiments.

\subsection{Simulated data and astrometric solution}

The key ingredients of the simulation process are a true astrometric catalogue, and a procedure for generating the observations and measurement uncertainties. For the true catalogue, we simply generated a set of stars uniformly distributed over the sky. In accordance with the theoretical treatment in Sect.~\ref{ss:observations}, we considered only along-scan measurements. For each star, we calculated the along-scan angle $\varphi$ at the moment when it crosses the middle of the relevant FOV. We, of course, do not include in our simulations' numerous complexities, which hamper real satellite observations such as attitude noise  and inadequacy of the calibration. These effects are modelled as random noise with formal uncertainty $\sigma_\varphi$ (specified below).  

The observations were simulated with the nominal \emph{Gaia} scanning law \citep{2016A&A...595A...1G}, which is a combination of three rotations (Fig.~\ref{fig:refframe}). First, the satellite rotates about the spin axis $\vec{z}$ with a period of 6 hours, that is the spin phase $\Omega$ linearly increased at the rate $\dot{\Omega}=60\arcsec\,\mathrm{s}^{-1}$. At the same time, the spin axis is in a slow precessional motion about the direction to the Sun completing 5.8 revolutions per year, with the solar aspect angle $\xi$ kept fixed. Finally, the Sun's motion along the ecliptic provides the coverage of the whole celestial sphere. The Sun advances 62\degr\ during one precession period. This `revolving scanning' produces a quasi-uniform temporal sampling and features a significant dependence on the ecliptic latitude.

Analysis of the post-fit residuals of the along-scan measurements in the EDR3 astrometric solution indicated that they are smallest for bright stars with $9\la G\la 13$ \citep[Appendix A in][]{2021A&A...649A...2L}. Whereas the residuals range from 100 to 200\,$\mu$as, the formal photon-statistical errors are between 50 and 70\,$\mu$as for these stars. Attributed to unmodelled errors, this difference will decrease in future data releases due to improved calibration. This is the best current estimation of the accuracy of the along-scan measurements and we adopted $\sigma_\varphi=100\,\mu\mathrm{as}$ as a representative value of the uncertainty. Furthermore, both the residuals and formal errors in EDR3 are weakly dependent on magnitude for $9\la G\la 13$. We, therefore, ignored the difference in stellar brightness and used the same uncertainty in all  simulated observations.

\begin{figure}
 \resizebox{\hsize}{!}{\includegraphics{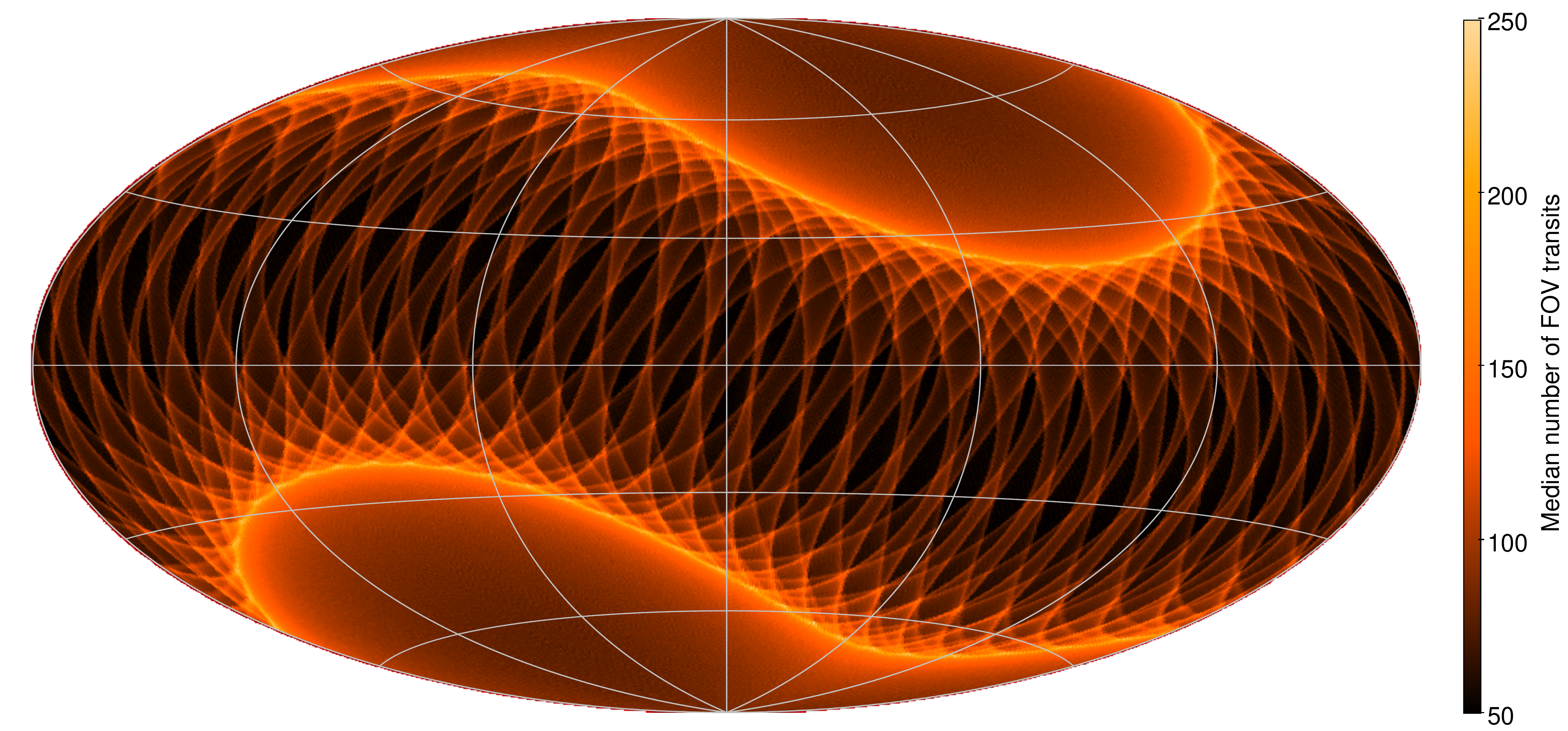}}
 \caption{Number of FOV transits per source in the equatorial coordinates for simulations covering 5 years of satellite operation. The distribution displays the essential characteristics of the \emph{Gaia}  scanning law, such as the rough symmetry around the ecliptic plane and high concentration of observations at ecliptic latitudes $\pm\left(90\degr-\xi\right)=\pm 45\degr$.}
 \label{fig:transits}
\end{figure}
 
The simulation process is straightforward. We assumed a Gaia-like FOV geometry: FOVs are rectangular with a nominal across- and along-scan size of 0.7\degr\ and 1\degr, respectively. The along-scan size slightly exceeded the \emph{Gaia} value of 0.7\degr; this was done only for convenience of computation and does not affect the results. The scanning law provided positions and orientations of the FOVs on the celestial sphere at any instance of time $t_l$. For each star fitting inside a FOV, its true instantaneous position was calculated using the \emph{Gaia} astrometric model \citep{lindegrenAGIS2012}. To account for the deviation from GR, light deflection was calculated with $\gamma-1=10^{-5}$. Finally, a transformation to a SRS provided a true value of the along-scan angle that was then perturbed with a Gaussian error with variance $\sigma_\varphi$ to simulate the observed value $\varphi_l$. This procedure generated, for each star, a set of pairs $\left(t_l,\varphi_l\right)$, which were treated as measurements and served as input data for the  subsequent astrometric solution.

In general, such an astrometric solution is a computationally heavy task, and it requires high-performance computing facilities, such as in the case of \emph{Gaia} \citep{lindegrenAGIS2012,Vecchiato+2018}. Such a computational complexity is mainly due to the need of solving for astrometric, attitude, and calibration unknowns at the same time. For our purposes, however, we need to solve the astrometric sphere for just the five conventional astrometric parameters (positions, parallax, and proper motions) for each star. Such a task corresponds to the solution of only the astrometric block of the \emph{Gaia} Astrometric Global Iterative Solution \citep{lindegrenAGIS2012}, namely a linear system of equations whose normal matrix is block-diagonal, with each block corresponding to the $5\times 5$ normal matrix of the astrometric parameters of a single star. The system is therefore embarrassingly parallel, which makes the solution much easier.
  
\subsection{Small-scale solutions}
\label{ss:small}

\begin{figure}
 \resizebox{\hsize}{!}{\includegraphics{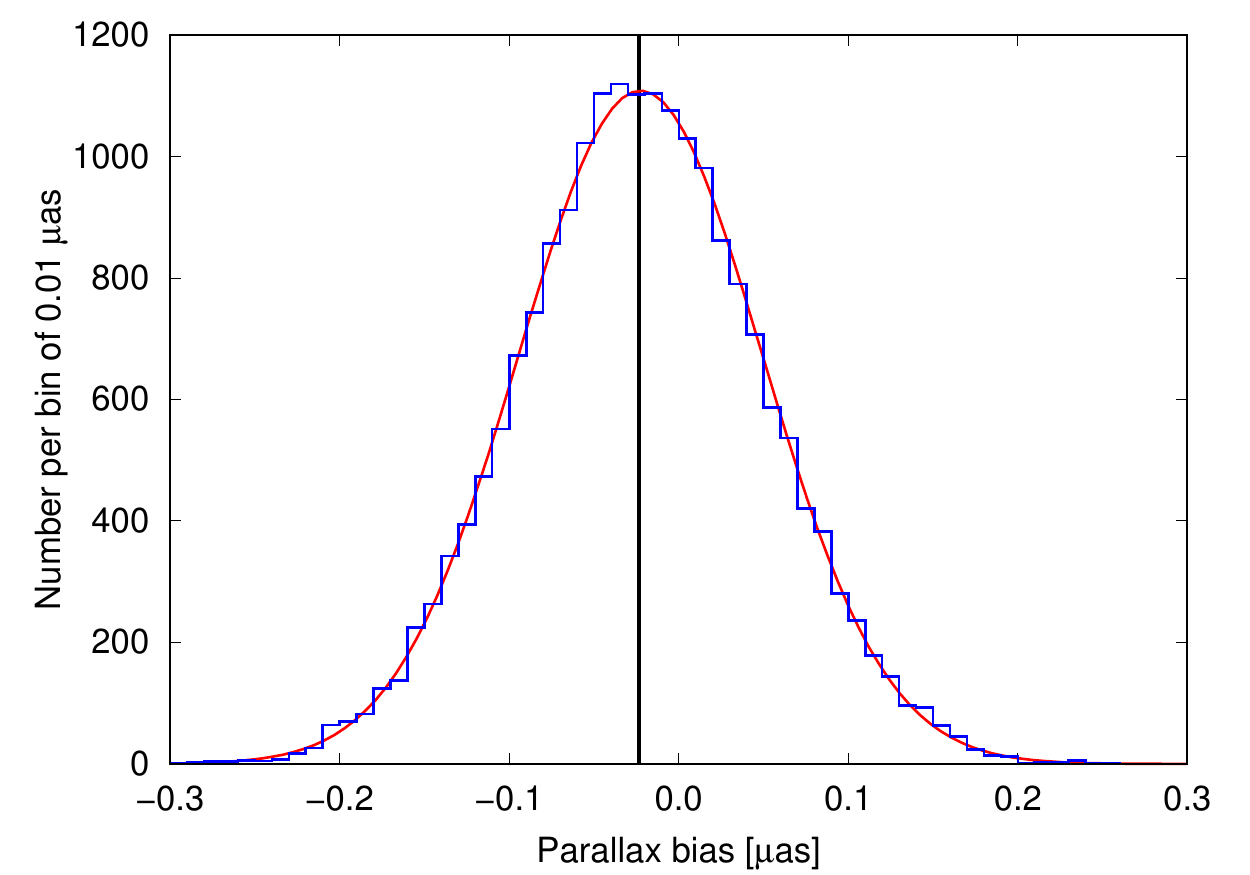}}
 \caption{Histogram of the parallax zero point found in 20\,000 astrometric solutions for $10^5$ stars using simulated observations with $\gamma-1=10^{-5}$. The distribution has a mean of $-0.023$\ $\mu$as, shown with the vertical solid line, while its standard deviation is 0.072\ $\mu$as. The red curve is a Gaussian distribution with the same parameters.}
 \label{fig:bias_par}
\end{figure}

As first test, we considered 20\,000 astrometric solutions, each containing 100\,000 stars uniformly distributed over the sky. The observations covered 5 years with no dead time. Figure~\ref{fig:transits} shows the sky map of the observation density. 
For each solution, we calculated the global parallax shift, defined as the average difference between the computed and true parallaxes. Thus, this Monte Carlo experiment provided 20\,000 estimates of the parallax zero point, whose distribution is shown in Fig.~\ref{fig:bias_par}. The mean zero point is $-0.023$\ $\mu$as and the standard deviation is 0.072\ $\mu$as. The mean agrees well with the theoretical prediction given by Eq.~(\ref{eq:parShift}) and we now demonstrate that the scatter is also in agreement with the expected outcome of this kind of experiment.
 
Equation~(\ref{eq:dpar_general}) is valid for an arbitrary variation in the along-scan instrument angle, including both systematic and random ones.  Above, we considered the systematic variations due to a change in PPN $\gamma$. With random errors included, the instrument angle variation is modelled as the sum of two components,
\begin{equation}\label{eq:dphi_i}
        \delta\varphi_l=g_l\,\delta\gamma+\varepsilon_l
.\end{equation}
The noise term $\varepsilon_l$ is assumed to be a centred, uncorrelated random variable with standard deviation $\sigma_\varphi$.
Substituting this formula in the general Eq.~(\ref{eq:dpar_general}), we find a change in the parallax of a given star, designated by the index $i$,
\begin{equation}
        \delta\varpi_i=\delta\varpi_\gamma-\frac{2}{R\sin\xi}\frac{1}{n_i}\sum_{l=1}^{n_i}\varepsilon_l\sin\theta_l\,,
\end{equation}
where $\delta\varpi_\gamma$ is the parallax shift induced by a variation in PPN $\gamma$, given by Eq.~(\ref{eq:parShift}), and $n_i$ is the number of observations of star $i$.

Averaging individual parallax errors gives the global parallax offset for a specific solution 
\begin{equation}\label{eq:Deltapi}
        \Delta\varpi=\frac{1}{N}\sum_{i=1}^N\delta\varpi_i=\delta\varpi_\gamma-\frac{2}{R\sin\xi}\frac{1}{N}\sum_{i=1}^N\frac{1}{n_i}\sum_{l=1}^{n_i}\varepsilon_l\sin\theta_l\,,
\end{equation}
where $N$ is the total number of stars in the solution. In our simulations, the observations are statistically independent. Noticing that 
$\left\langle\varepsilon^2\right\rangle=\sigma_\varphi^2$ and 
$\left\langle\sin^2\theta\right\rangle=1/2$, we therefore found the estimation of the standard deviation in the parallax zero point
\begin{equation}\label{eq:sigmaDeltaPi}
        \sigma_{\Delta\varpi}=\frac{\sigma_\varphi}{R\sin\xi}\sqrt{\frac{2\left\langle 1/n\right\rangle}{N}}\,,
\end{equation}
where $\left\langle 1/n\right\rangle=\left(1/N\right)\sum_{i=1}^N\left(1/n_i\right)$ is the harmonic mean of the number of observations per star; $\left\langle 1/n\right\rangle=0.012$ for the 5 yr mission.

\begin{figure}
 \resizebox{\hsize}{!}{\includegraphics{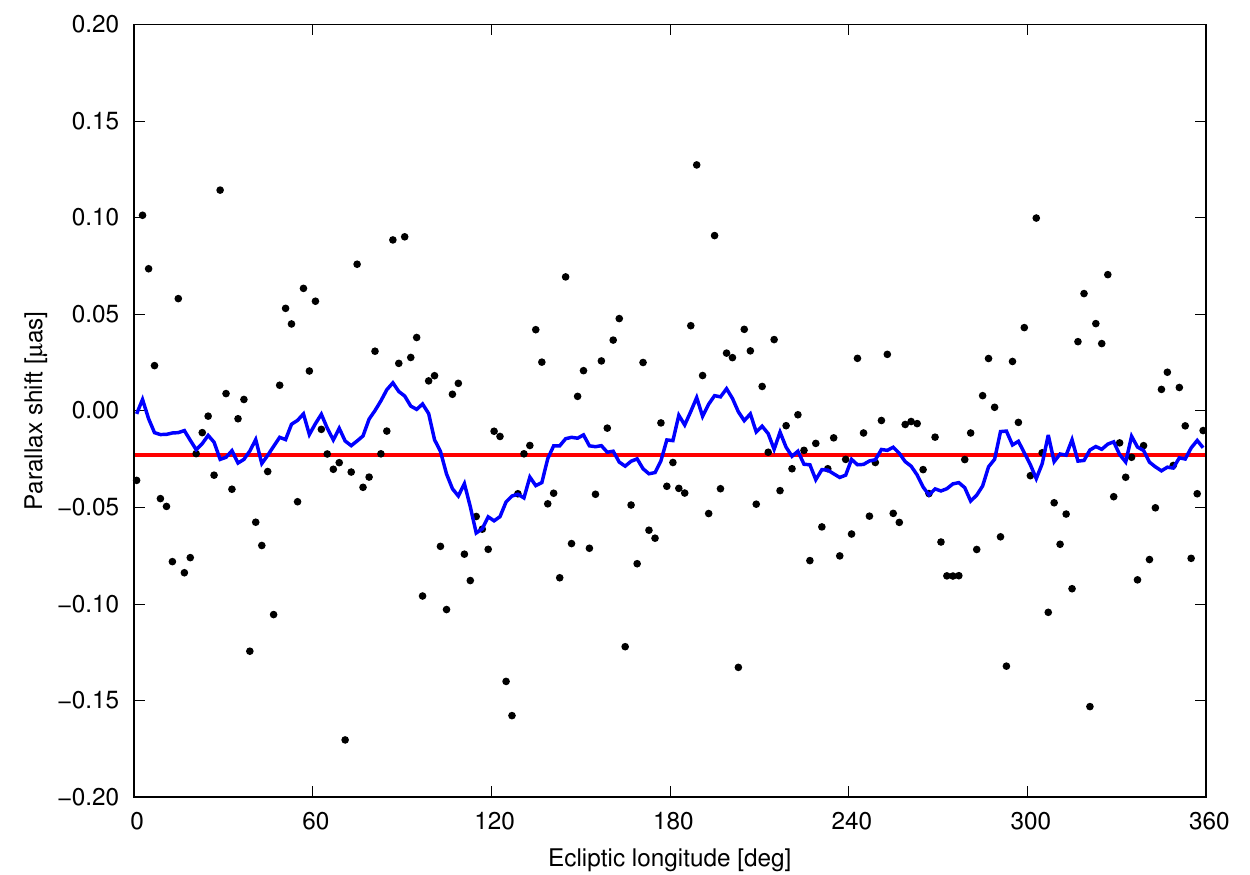}}
 \caption{Mean parallax shift, in two-degree bins, versus ecliptic longitude in the solution with 14 million stars for a mission length of 10 years. The blue curve is the running mean smoothed with a linear kernel of width 13. The theoretical value of $-0.023$\ $\mu$as is shown with the red line.}
 \label{fig:ecllon}
\end{figure}

For the \emph{Gaia} parameters, Eq.~(\ref{eq:sigmaDeltaPi}) gives $\sigma_{\Delta\varpi}=0.069\ \mu\mathrm{as}$, which is in agreement with the experimental value of 0.072\ $\mu$as.  Figure~\ref{fig:bias_par} shows that the distribution of the parallax shift mainly follows a normal one. The slight  excess on the left to the peak is attributable to various small effects, such as the finite FOV size and weak dependence of $R$ and $\xi$ on time, as well as to the scanning law asymmetry discussed in Sect.~\ref{sec:asymmetry}. Thus, the results of this Monte Carlo experiment confirm the theoretical expectations for the basic statistical properties of the  parallax zero point.

\subsection{Large-scale solution}

The small-scale simulation, considered in Sect.~\ref{ss:small}, allows us to examine the parallax zero point, only in a statistical sense, using a large number of independent solutions. We now aim to study the impact of a PPN $\gamma$ variation on the zero point by conducting one large-scale astrometric solution with better  accuracy for a vastly increased number of stars. For a fixed measurement uncertainty, the overall accuracy of a solution can be improved by extending the time coverage. To this end, we considered a mission that was twice as long, lasting 10 yr. The formal uncertainties of the parallaxes scale with the observation duration as $T^{-1/2}$ \citep{2021A&A...649A...2L}. Thus, a twofold increase in the mission length results in an improvement in the accuracy by a factor 0.71 for the parallaxes. The 10-yr duration corresponds to the foreseen \emph{Gaia} extended mission lifetime \citep{Gaia-CollaborationEDR32020}. Accordingly, we conducted a solution with 14 million stars, which is roughly equal to the number of primary stars in the Gaia EDR3.

A vastly increased number of observations brought substantial improvement for stability of the parallax bias estimation against noise. The parallax zero point of  $-0.020$\ $\mu$as was detected in the large-scale astrometric solution. The median parallax uncertainty was found to be 17\ $\mu$as, and therefore the expected precision of the zero point determination is $\left(17\ \mu\mathrm{as}\right)\times[14\times10^6]^{-1/2}\simeq4.5\times10^{-3}\ \mu\mathrm{as}$. Thus,  the detected value of the zero point agrees with the theoretical prediction of $-0.023$\ $\mu$as to within one standard deviation. There is, however, a considerable scatter in the data, as illustrated in Fig.~\ref{fig:ecllon}. The points show a mean parallax shift, computed in bins of two degrees, as a function of ecliptic latitude. The bins contain 78\,000 stars on average. The values in the bins range from $-0.17$ to $0.12$\ $\mu$as.

We carried out an additional test to examine the consistency of the parallax zero point estimation: all of the data were split into odd- and even-numbered stars, resulting in two subsets of 7 million stars each. Independent astrometric solutions were calculated for the two data sets and the mean parallax shifts were found to be $-0.025$ and $-0.016$\ $\mu$as. The formal uncertainty is $6.4\times10^{-3}\ \mu\mathrm{as}$ for $7\times10^6$ stars, and therefore these results also confirm the theoretical prediction at a 1-$\sigma$ level. The good agreement between the theoretical expectation and the experimental results demonstrates the validity of the mathematical treatment
and correct functioning of the simulation procedure.  

\begin{figure}
 \resizebox{\hsize}{!}{\includegraphics{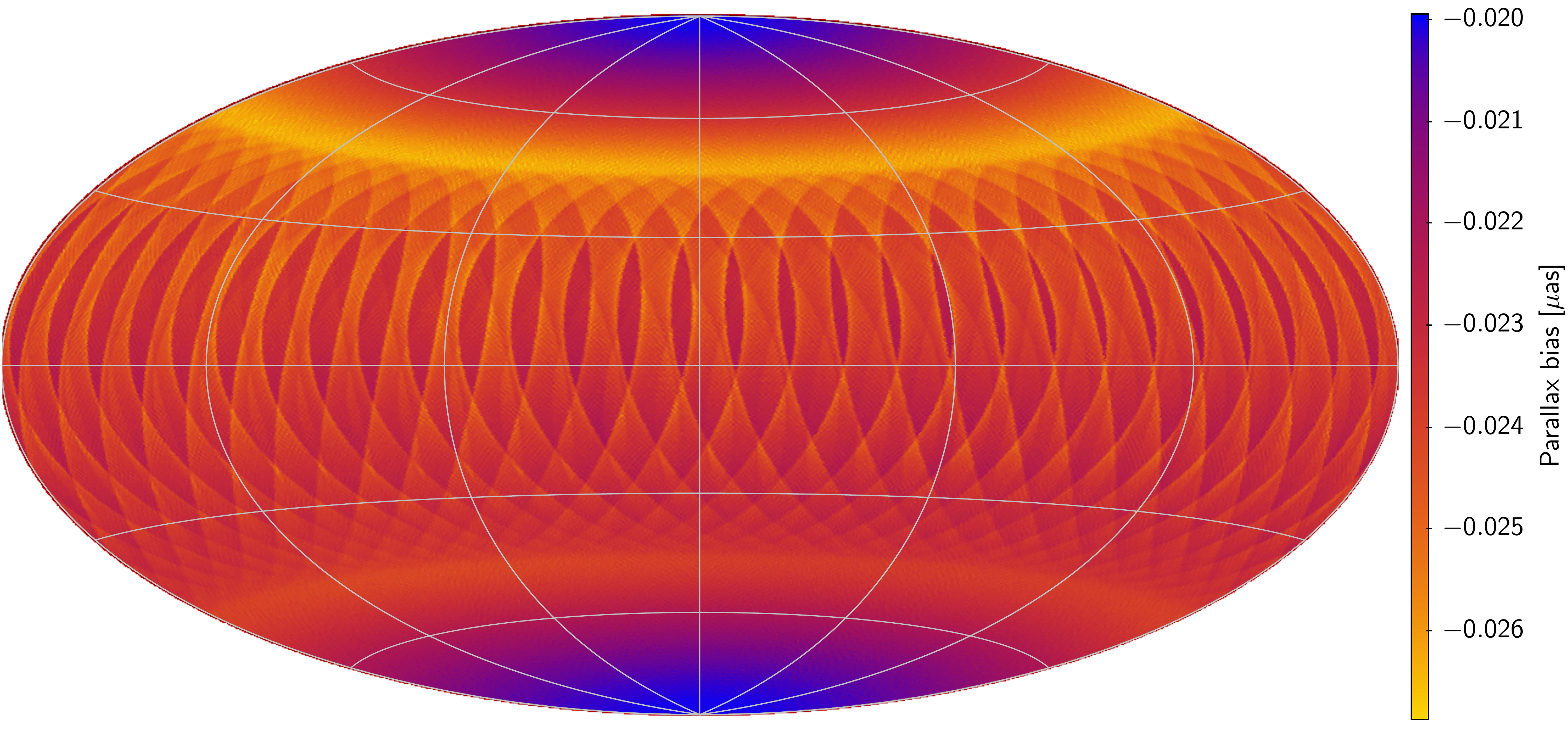}}
 \caption{Map of the parallax error in ecliptic coordinates obtained in the astrometric solution using noise-free observations of five million stars for an extended 10 yr mission with the nominal \emph{Gaia} scanning law.}
 \label{fig:ecliptic2d}
\end{figure}

\section{Dependence of parallax bias on star position}
\label{sec:asymmetry}

We now turn to a study of how the parallax bias varies with the distance of a star from the ecliptic. This kind of systematics presumably exists by virtue of the strong dependence of the scanning law on ecliptic latitude (Fig.~\ref{fig:transits}). Providing clear evidence for the parallax zero point due to a variation in the PPN, the large-scale solution does not enable us to study the finer details of the bias because of a considerable scatter among individual stars. Even the largely smoothed data, drawn in Fig.~\ref{fig:ecllon} with the blue curve, are dominated by random variations around the mean. 

In our numerical tests, with the GR violation parametrised as $\gamma-1=10^{-5}$ and the current assessment of the Gaia observation uncertainty of 100\ $\mu$as, the mean value of the parallax bias, $-0.023$\ $\mu$as, is three orders of magnitude less than the typical accuracy of individual parallaxes, being 17\ $\mu$as for a 10\ yr mission. In the next simulation run, we considered five million stars randomly distributed on the sky. With the average star density of 121 deg$^{-2}$, these sources constitute a quasi-uniform grid dense enough to accurately map the parallax bias over the entire celestial sphere. The generation of measurement noise was not activated in this simulation to yield noiseless observations. 

\begin{figure}
 \resizebox{\hsize}{!}{\includegraphics{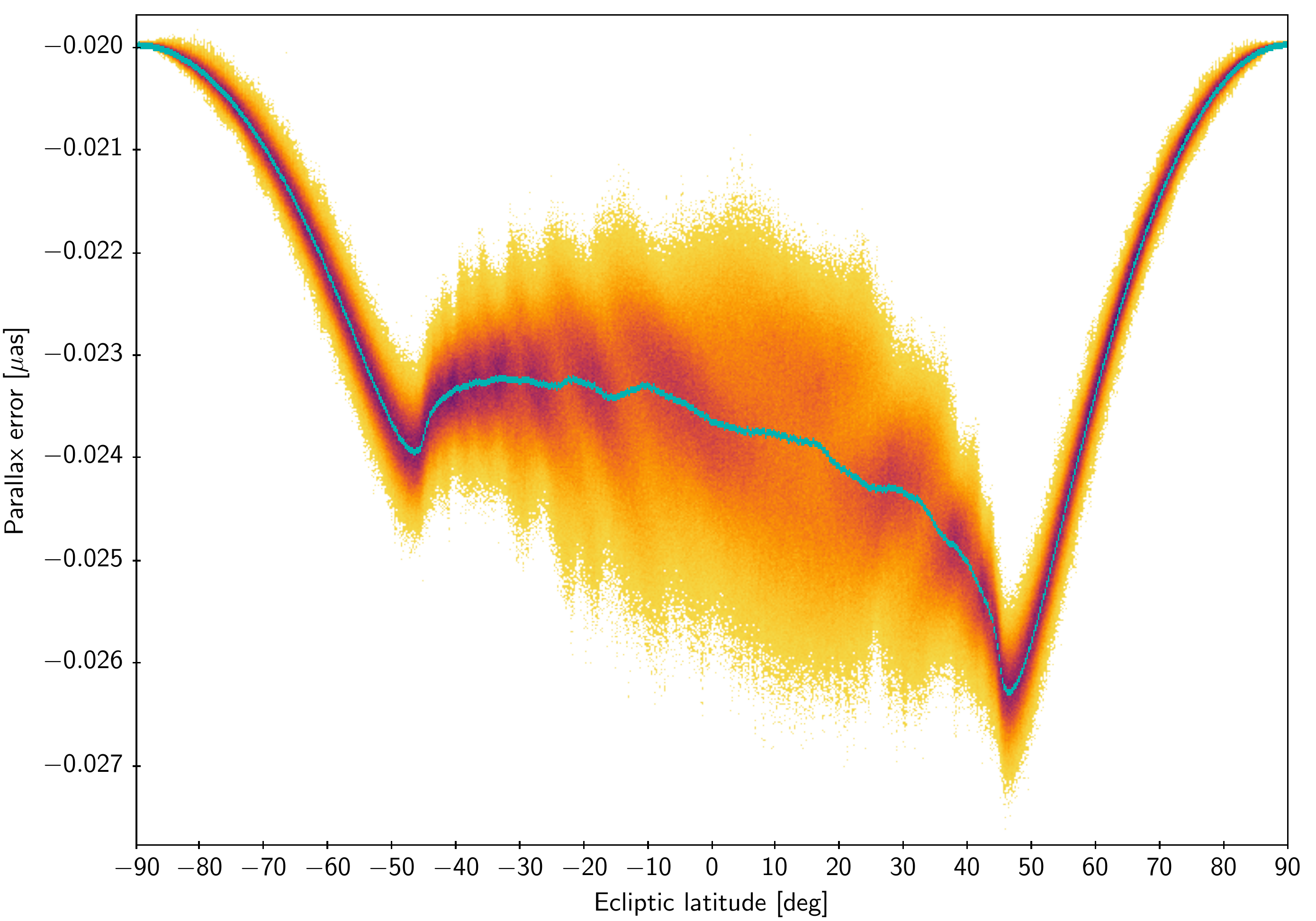}}
 \caption{Parallax error plotted against the ecliptic latitude in the astrometric solution with five million stars with the nominal \emph{Gaia} scanning law. 
 The cyan curve is the running median.}
 \label{fig:ecliptic}
\end{figure}

The all-sky map of the resulting parallax error is drawn in Fig.~\ref{fig:ecliptic2d}. The mean of this distribution is $-0.024$\ $\mu$as, with a dispersion of $0.001$\ $\mu$as; this bias confirms the theoretical expectations. Although the appearance of the parallax error generally bears the pattern of the revolving scanning, it has a noteworthy feature: the north-south asymmetry in ecliptic coordinates. The absolute value of the bias in the northern ecliptic hemisphere is systematically larger compared to the southern one. This phenomenon is further illustrated in Fig.~\ref{fig:ecliptic}, where the dependence of the parallax error on the ecliptic latitude is shown. The asymmetry is clearly seen in this plot.

The scanning law provides the sky coverage, which is geometrically symmetric with respect to the ecliptic equator. As exemplified in Fig.~\ref{fig:transits}, the scanning density does not exhibit a systematic dependence on the ecliptic latitude, apart from the slight non-uniformity in the ecliptic belt, that is for $\left|\beta\right|\la45\degr$, due to an interaction of the solar orbital motion and the spin axis precession. It is, therefore, natural to assume that the parallax bias possesses the same symmetry. The numerical experiments, however, make us call this expectation into question. We argue below that such an intricate behaviour of the bias results from an asymmetry inherent in the revolving scanning.

The critical factor in both the parallactic effect and gravitational light deflection is the angle between directions to the star and the Sun. We examined the distribution of this angle and found that for stars in the northern ecliptical polar cap $\left(\beta\ga46\degr\right),$ its mean value is 87\fdg3; whereas, in the southern cap $\left(\beta\la-46\degr\right),$ the mean is 92\fdg7.
Thus, the stars in the northern cap are mainly observed closer to the Sun, compared to those in the southern cap.

The reason for that can be the following. In the nominal \emph{Gaia} scanning law, the spin axis describes the loops is the ecliptic belt. The spin axis rotates counterclockwise, as seen from the satellite \citep{2016A&A...595A...1G}. As a result, the direction of the precession motion coincides with the direction of the Sun for $\beta<0$, whereas for positive $\beta$ these directions are the opposite. This means that the axis  covers a longer path in the southern ecliptic hemisphere. Corresponding scanning circles cover a larger area and consequently contain more stars.   

For this reason, stars in the northern hemisphere are observed closer than 90\degr\ to the Sun on average, while the stars in the southern hemisphere are observed further than 90\degr\ away from the Sun on average. The parallactic effect is symmetric with respect to 90\degr. On the contrary, the light deflection rises as the angle to the Sun decreases. In the north, this angle is systematically smaller than 90\degr, while in the south it is systematically larger than 90\degr. Consequently, the light deflection is stronger for the stars in the northern polar cap for a Gaia observation. As a result, the detected parallax shift is also larger in the north.

\begin{figure}
 \resizebox{\hsize}{!}{\includegraphics{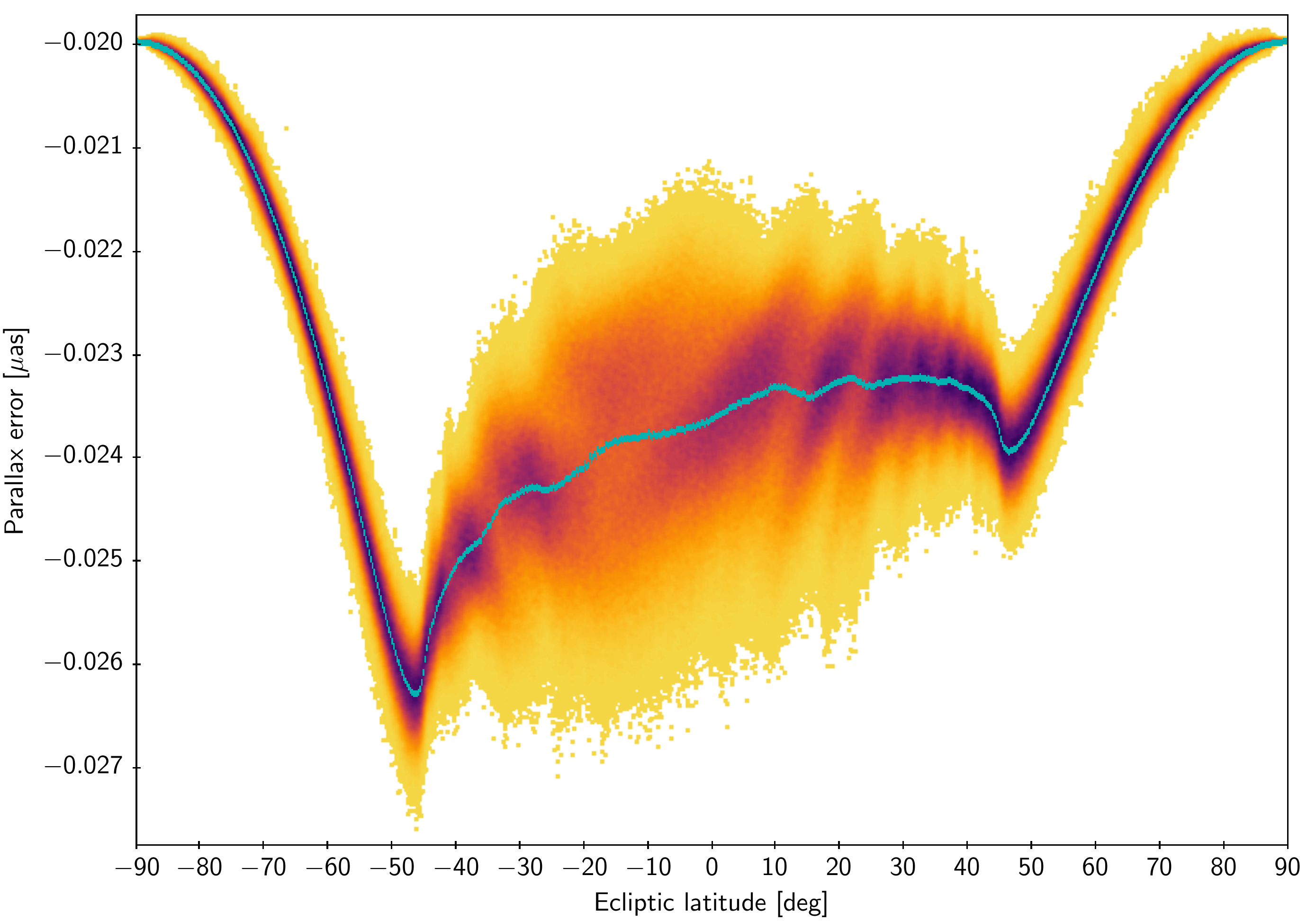}}
 \caption{As Fig.~\ref{fig:ecliptic}, but for the \emph{Gaia} scanning law with a reversed direction for the precession, with the spin axis revolving around the Sun clockwise, as seen from the satellite.}
 \label{fig:eclipticrev}
\end{figure}

To verify this hypothesis, an additional simulation was performed with a reversed motion of the spin axis. In this case, the spin axis spends more time in the northern ecliptic hemisphere. The dependence of the parallax bias on the ecliptic latitude found in the experiment with the reversed precession is displayed in Fig.~\ref{fig:eclipticrev}. Comparison of Figs.~\ref{fig:ecliptic} and \ref{fig:eclipticrev} shows that the changes of the precession direction have an inverse effect on the parallax bias. This provides evidence for the adequacy of the proposed explanation for the asymmetry in the parallax bias.

\section{Discussion and conclusions}
\label{sec:conclusions}

In this paper we investigated the consequences of the well-known correlation between parallaxes and the PPN $\gamma$ parameter in astrometric observables, with an emphasis on the impact of a bias of the PPN $\gamma$ parameter on the resulting parallax uncertainties in \emph{Gaia}-like missions.
This systematic error translates into a corresponding bias for parallaxes, known as the zero-point parallax error, and the existence of such an issue in the current \emph{Gaia} solutions was one of the motivations that inspired our investigation.

A straightforward statistical analysis provided a relation, never before mentioned in the literature, between the zero-point parallax error and the PPN $\gamma$ bias for \emph{Gaia}-like missions, namely for observations based on a scanning strategy. Such a relation was then numerically verified with a series of Monte Carlo runs of a simulated \emph{Gaia} mission. This proved that the size of the zero-point parallax error observed in the actual \emph{Gaia} solution cannot be attributed to the effect of a possible deviation in the PPN $\gamma$ parameter from the value assumed in GR, which has been adopted by all of the current astrometric models in the reduction pipelines. Moreover, it also implies that any bias of PPN $\gamma$ within the known experimental uncertainty cannot generate a systematic effect on the parallaxes beyond the target accuracy of this satellite, and therefore it is negligible for this mission. Nonetheless, this remains potentially relevant for future similar missions with the precision goal being improved by one order of magnitude.

In this respect, when we analysed the parallax bias effect on different regions of the sky, we discovered an interesting, but qualitatively expected, dependence on the ecliptic latitude that we dubbed `diavoletto' (`little devil' in Italian). Such a dependence is naturally introduced by the scanning law, and thus has to be observed in every mission sharing this type of observing strategy, and that depends on the precession direction direction of precession of the spin axis of the satellite in a significant manner.

As mentioned in \citet{2021A&A...649A...2L}, the same precession direction of the spin axis that produces the `diavoletto' is also the very origin of a strong correlation between the AL parallax factor and the AC rate. The cited paper stresses that this correlation can bias the parallaxes of stars whose image parameter determination is significantly impacted by the resulting AC smearing. This effect was deemed important enough to reverse the precession direction of the satellite for about one year in order to break this correlation.

This calls for further investigation in this field, as it is thus quite possible that these are just two examples of a whole set of previously  correlations, whose effects on the \emph{Gaia} and \emph{Gaia}-like missions are still not known. 

Following this line of reasoning, we may venture into a more general extrapolation: even a pointed mission, not using any kind of scanning law, is still susceptible to such an effect whenever the sampling scheme includes any asymmetry among relevant regions of the sky.  Even so, the systematic error can be estimated for relevant cases, and taken care of in data reduction, if there is sufficient confidence on the $\gamma$ parameter. A more reliable approach would of course consist in mitigating the effect at its origin by enforcing suitable `sampling rules'. By comparison of Figs.~\ref{fig:ecliptic} and \ref{fig:eclipticrev}, a reversed scanning law applied for $50\%$ of the mission lifetime would effectively compensate for the systematic error by equalising the exposure time excess between hemispheres and still provide adequate observations for the purposes of most astrometric and astrophysical measurements. Optimal strategies may be devised by means of dedicated studies in order to improve on efficiency with respect to satellite operation and/or specific science cases.

\begin{acknowledgements}
This work was supported by the Italian Space Agency ASI under contract 2018-24-HH.0 in support of the Italian participation to the Gaia mission, 
and by the grant Astrometric Science and Technology Roadmap for Astrophysics (ASTRA) 
from the Italian Ministry of Foreign Affairs and International Cooperation (MAECI). 
The numerical simulations were executed at the CINECA supercomputing centre.
A.G.B. is thankful to the Astrophysical Observatory of Torino for their warm hospitality during his short-term visits.
A.G.B. is also grateful to Lennart Lindegren (Lund Observatory) for kind permission to use his drawing software. 
This research made use of NASA's Astrophysics Data System. Diagrams were produced using the astronomy data visualisation software TOPCAT \citep{Taylor2005}.

\end{acknowledgements}

\bibliographystyle{aa}
\bibliography{parGamma1}

\begin{thebibliography}{18}
\expandafter\ifx\csname natexlab\endcsname\relax\def\natexlab#1{#1}\fi

\bibitem[{{Bertotti} {et~al.}(2003){Bertotti}, {Iess}, \&
  {Tortora}}]{Bertotti+2003}
{Bertotti}, B., {Iess}, L., \& {Tortora}, P. 2003, \nat, 425, 374

\bibitem[{{Butkevich} {et~al.}(2017){Butkevich}, {Klioner}, {Lindegren},
  {Hobbs}, \& {van Leeuwen}}]{Butkevich+2017}
{Butkevich}, A.~G., {Klioner}, S.~A., {Lindegren}, L., {Hobbs}, D., \& {van
  Leeuwen}, F. 2017, \aap, 603, A45

\bibitem[{{ESA}(1997)}]{1997ESASP1200.....E}
{ESA}, ed. 1997, ESA Special Publication, Vol. 1200, {The HIPPARCOS and TYCHO
  catalogues. Astrometric and photometric star catalogues derived from the ESA
  HIPPARCOS Space Astrometry Mission}

\bibitem[{Froeschl\'e {et~al.}(1997)Froeschl\'e, Mignard, \&
  Arenou}]{Froeschle+1997}
Froeschl\'e, M., Mignard, F., \& Arenou, F. 1997, in Hipparcos-Venice `97
  Proceedings, ESA SP-402, 49

\bibitem[{{Gaia Collaboration} {et~al.}(2021){Gaia Collaboration}, {Brown},
  {Vallenari}, {Prusti}, {de Bruijne}, {Babusiaux}, \&
  {Biermann}}]{Gaia-CollaborationEDR32020}
{Gaia Collaboration}, {Brown}, A.~G.~A., {Vallenari}, A., {et~al.} 2021, A\&A,
  649, A1

\bibitem[{{Gaia Collaboration} {et~al.}(2016){Gaia Collaboration}, {Prusti},
  {de Bruijne}, {Brown}, {Vallenari}, {Babusiaux}, {Bailer-Jones}, {Bastian},
  {Biermann}, {Evans}, {Eyer}, {Jansen}, {Jordi}, {Klioner}, {Lammers},
  {Lindegren}, {Luri}, {Mignard}, {Milligan}, {Panem}, {Poinsignon},
  {Pourbaix}, {Randich}, {Sarri}, {Sartoretti}, {Siddiqui}, {Soubiran},
  {Valette}, {van Leeuwen}, {Walton}, {Aerts}, {Arenou}, {Cropper}, {Drimmel},
  {H{\o}g}, {Katz}, {Lattanzi}, {O'Mullane}, {Grebel}, {Holland}, {Huc},
  {Passot}, {Bramante}, {Cacciari}, {Casta{\~n}eda}, {Chaoul}, {Cheek}, {De
  Angeli}, {Fabricius}, {Guerra}, {Hern{\'a}ndez}, {Jean-Antoine-Piccolo},
  {Masana}, {Messineo}, {Mowlavi}, {Nienartowicz}, {Ord{\'o}{\~n}ez-Blanco},
  {Panuzzo}, {Portell}, {Richards}, {Riello}, {Seabroke}, {Tanga},
  {Th{\'e}venin}, {Torra}, {Els}, {Gracia-Abril}, {Comoretto},
  {Garcia-Reinaldos}, {Lock}, {Mercier}, {Altmann}, {Andrae}, {Astraatmadja},
  {Bellas-Velidis}, {Benson}, {Berthier}, {Blomme}, {Busso}, {Carry},
  {Cellino}, {Clementini}, {Cowell}, {Creevey}, {Cuypers}, {Davidson}, {De
  Ridder}, {de Torres}, {Delchambre}, {Dell'Oro}, {Ducourant}, {Fr{\'e}mat},
  {Garc{\'\i}a-Torres}, {Gosset}, {Halbwachs}, {Hambly}, {Harrison}, {Hauser},
  {Hestroffer}, {Hodgkin}, {Huckle}, {Hutton}, {Jasniewicz}, {Jordan},
  {Kontizas}, {Korn}, {Lanzafame}, {Manteiga}, {Moitinho}, {Muinonen},
  {Osinde}, {Pancino}, {Pauwels}, {Petit}, {Recio-Blanco}, {Robin}, {Sarro},
  {Siopis}, {Smith}, {Smith}, {Sozzetti}, {Thuillot}, {van Reeven}, {Viala},
  {Abbas}, {Abreu Aramburu}, {Accart}, {Aguado}, {Allan}, {Allasia},
  {Altavilla}, {{\'A}lvarez}, {Alves}, {Anderson}, {Andrei}, {Anglada Varela},
  {Antiche}, {Antoja}, {Ant{\'o}n}, {Arcay}, {Atzei}, {Ayache}, {Bach},
  {Baker}, {Balaguer-N{\'u}{\~n}ez}, {Barache}, {Barata}, {Barbier}, {Barblan},
  {Baroni}, {Barrado y Navascu{\'e}s}, {Barros}, {Barstow}, {Becciani},
  {Bellazzini}, {Bellei}, {Bello Garc{\'\i}a}, {Belokurov}, {Bendjoya},
  {Berihuete}, {Bianchi}, {Bienaym{\'e}}, {Billebaud}, {Blagorodnova},
  {Blanco-Cuaresma}, {Boch}, {Bombrun}, {Borrachero}, {Bouquillon}, {Bourda},
  {Bouy}, {Bragaglia}, {Breddels}, {Brouillet}, {Br{\"u}semeister},
  {Bucciarelli}, {Budnik}, {Burgess}, {Burgon}, {Burlacu}, {Busonero}, {Buzzi},
  {Caffau}, {Cambras}, {Campbell}, {Cancelliere}, {Cantat-Gaudin}, {Carlucci},
  {Carrasco}, {Castellani}, {Charlot}, {Charnas}, {Charvet}, {Chassat},
  {Chiavassa}, {Clotet}, {Cocozza}, {Collins}, {Collins}, {Costigan}, {Crifo},
  {Cross}, {Crosta}, {Crowley}, {Dafonte}, {Damerdji}, {Dapergolas}, {David},
  {David}, {De Cat}, {de Felice}, {de Laverny}, {De Luise}, {De March}, {de
  Martino}, {de Souza}, {Debosscher}, {del Pozo}, {Delbo}, {Delgado},
  {Delgado}, {di Marco}, {Di Matteo}, {Diakite}, {Distefano}, {Dolding}, {Dos
  Anjos}, {Drazinos}, {Dur{\'a}n}, {Dzigan}, {Ecale}, {Edvardsson}, {Enke},
  {Erdmann}, {Escolar}, {Espina}, {Evans}, {Eynard Bontemps}, {Fabre},
  {Fabrizio}, {Faigler}, {Falc{\~a}o}, {Farr{\`a}s Casas}, {Faye}, {Federici},
  {Fedorets}, {Fern{\'a}ndez-Hern{\'a}ndez}, {Fernique}, {Fienga}, {Figueras},
  {Filippi}, {Findeisen}, {Fonti}, {Fouesneau}, {Fraile}, {Fraser}, {Fuchs},
  {Furnell}, {Gai}, {Galleti}, {Galluccio}, {Garabato}, {Garc{\'\i}a-Sedano},
  {Gar{\'e}}, {Garofalo}, {Garralda}, {Gavras}, {Gerssen}, {Geyer}, {Gilmore},
  {Girona}, {Giuffrida}, {Gomes}, {Gonz{\'a}lez-Marcos},
  {Gonz{\'a}lez-N{\'u}{\~n}ez}, {Gonz{\'a}lez-Vidal}, {Granvik}, {Guerrier},
  {Guillout}, {Guiraud}, {G{\'u}rpide}, {Guti{\'e}rrez-S{\'a}nchez}, {Guy},
  {Haigron}, {Hatzidimitriou}, {Haywood}, {Heiter}, {Helmi}, {Hobbs},
  {Hofmann}, {Holl}, {Holland }, {Hunt}, {Hypki}, {Icardi}, {Irwin}, {Jevardat
  de Fombelle}, {Jofr{\'e}}, {Jonker}, {Jorissen}, {Julbe}, {Karampelas},
  {Kochoska}, {Kohley}, {Kolenberg}, {Kontizas}, {Koposov}, {Kordopatis},
  {Koubsky}, {Kowalczyk}, {Krone-Martins}, {Kudryashova}, {Kull}, {Bachchan},
  {Lacoste-Seris}, {Lanza}, {Lavigne}, {Le Poncin-Lafitte}, {Lebreton},
  {Lebzelter}, {Leccia}, {Leclerc}, {Lecoeur-Taibi}, {Lemaitre}, {Lenhardt},
  {Leroux}, {Liao}, {Licata}, {Lindstr{\o}m}, {Lister}, {Livanou}, {Lobel},
  {L{\"o}ffler}, {L{\'o}pez}, {Lopez-Lozano}, {Lorenz}, {Loureiro},
  {MacDonald}, {Magalh{\~a}es Fernandes}, {Managau}, {Mann}, {Mantelet},
  {Marchal}, {Marchant}, {Marconi}, {Marie}, {Marinoni}, {Marrese},
  {Marschalk{\'o}}, {Marshall}, {Mart{\'\i}n-Fleitas}, {Martino}, {Mary},
  {Matijevi{\v{c}}}, {Mazeh}, {McMillan}, {Messina}, {Mestre}, {Michalik},
  {Millar}, {Miranda}, {Molina}, {Molinaro}, {Molinaro}, {Moln{\'a}r},
  {Moniez}, {Montegriffo}, {Monteiro}, {Mor}, {Mora}, {Morbidelli}, {Morel},
  {Morgenthaler}, {Morley}, {Morris}, {Mulone}, {Muraveva}, {Musella},
  {Narbonne}, {Nelemans}, {Nicastro}, {Noval}, {Ord{\'e}novic},
  {Ordieres-Mer{\'e}}, {Osborne}, {Pagani}, {Pagano}, {Pailler}, {Palacin},
  {Palaversa}, {Parsons}, {Paulsen}, {Pecoraro}, {Pedrosa}, {Pentik{\"a}inen},
  {Pereira}, {Pichon}, {Piersimoni}, {Pineau}, {Plachy}, {Plum}, {Poujoulet},
  {Pr{\v{s}}a}, {Pulone}, {Ragaini}, {Rago}, {Rambaux}, {Ramos-Lerate},
  {Ranalli}, {Rauw}, {Read}, {Regibo}, {Renk}, {Reyl{\'e}}, {Ribeiro},
  {Rimoldini}, {Ripepi}, {Riva}, {Rixon}, {Roelens}, {Romero-G{\'o}mez},
  {Rowell}, {Royer}, {Rudolph}, {Ruiz-Dern}, {Sadowski}, {Sagrist{\`a}
  Sell{\'e}s}, {Sahlmann}, {Salgado}, {Salguero}, {Sarasso}, {Savietto},
  {Schnorhk}, {Schultheis}, {Sciacca}, {Segol}, {Segovia}, {Segransan},
  {Serpell}, {Shih}, {Smareglia}, {Smart}, {Smith}, {Solano}, {Solitro},
  {Sordo}, {Soria Nieto}, {Souchay}, {Spagna}, {Spoto}, {Stampa}, {Steele},
  {Steidelm{\"u}ller}, {Stephenson}, {Stoev}, {Suess}, {S{\"u}veges}, {Surdej},
  {Szabados}, {Szegedi-Elek}, {Tapiador}, {Taris}, {Tauran}, {Taylor},
  {Teixeira}, {Terrett}, {Tingley}, {Trager}, {Turon}, {Ulla}, {Utrilla},
  {Valentini}, {van Elteren}, {Van Hemelryck}, {van Leeuwen}, {Varadi},
  {Vecchiato}, {Veljanoski}, {Via}, {Vicente}, {Vogt}, {Voss}, {Votruba},
  {Voutsinas}, {Walmsley}, {Weiler}, {Weingrill}, {Werner}, {Wevers},
  {Whitehead}, {Wyrzykowski}, {Yoldas}, {{\v{Z}}erjal}, {Zucker}, {Zurbach},
  {Zwitter}, {Alecu}, {Allen}, {Allende Prieto}, {Amorim},
  {Anglada-Escud{\'e}}, {Arsenijevic}, {Azaz}, {Balm}, {Beck}, {Bernstein},
  {Bigot}, {Bijaoui}, {Blasco}, {Bonfigli}, {Bono}, {Boudreault}, {Bressan},
  {Brown}, {Brunet}, {Bunclark}, {Buonanno}, {Butkevich}, {Carret}, {Carrion},
  {Chemin}, {Ch{\'e}reau}, {Corcione}, {Darmigny}, {de Boer}, {de Teodoro}, {de
  Zeeuw}, {Delle Luche}, {Domingues}, {Dubath}, {Fodor}, {Fr{\'e}zouls},
  {Fries}, {Fustes}, {Fyfe}, {Gallardo}, {Gallegos}, {Gardiol}, {Gebran},
  {Gomboc}, {G{\'o}mez}, {Grux}, {Gueguen}, {Heyrovsky}, {Hoar}, {Iannicola},
  {Isasi Parache}, {Janotto}, {Joliet}, {Jonckheere}, {Keil}, {Kim},
  {Klagyivik}, {Klar}, {Knude}, {Kochukhov}, {Kolka}, {Kos}, {Kutka}, {Lainey},
  {LeBouquin}, {Liu}, {Loreggia}, {Makarov}, {Marseille}, {Martayan},
  {Martinez-Rubi}, {Massart}, {Meynadier}, {Mignot}, {Munari}, {Nguyen},
  {Nordlander}, {Ocvirk}, {O'Flaherty}, {Olias Sanz}, {Ortiz}, {Osorio},
  {Oszkiewicz}, {Ouzounis}, {Palmer}, {Park}, {Pasquato}, {Peltzer}, {Peralta},
  {P{\'e}turaud}, {Pieniluoma}, {Pigozzi}, {Poels}, {Prat}, {Prod'homme},
  {Raison}, {Rebordao}, {Risquez}, {Rocca-Volmerange}, {Rosen}, {Ruiz-Fuertes},
  {Russo}, {Sembay}, {Serraller Vizcaino}, {Short}, {Siebert}, {Silva},
  {Sinachopoulos}, {Slezak}, {Soffel}, {Sosnowska}, {Strai{\v{z}}ys}, {ter
  Linden}, {Terrell}, {Theil}, {Tiede}, {Troisi}, {Tsalmantza}, {Tur},
  {Vaccari}, {Vachier}, {Valles}, {Van Hamme}, {Veltz}, {Virtanen}, {Wallut},
  {Wichmann}, {Wilkinson}, {Ziaeepour}, \& {Zschocke}}]{2016A&A...595A...1G}
{Gaia Collaboration}, {Prusti}, T., {de Bruijne}, J.~H.~J., {et~al.} 2016,
  \aap, 595, A1

\bibitem[{{Hobbs} {et~al.}(2009){Hobbs}, {Holl}, {Lindegren}, {Raison},
  {Klioner}, \& {Butkevich}}]{Hobbs+2009}
{Hobbs}, D., {Holl}, B., {Lindegren}, L., {et~al.} 2009, in Relativity in
  Fundamental Astronomy, IAU Symposium 261, ed. S.~A. {Klioner}, P.~K.
  {Seidelman}, \& M.~H. {Soffel}, 315

\bibitem[{Klioner(2003)}]{klioner2003}
Klioner, S. 2003, ApJ, 125, 1580

\bibitem[{{Lindegren} {et~al.}(2021{\natexlab{a}}){Lindegren}, {Bastian},
  {Biermann}, {Bombrun}, {de Torres}, {Gerlach}, {Geyer}, {Hern{\'a}ndez},
  {Hilger}, {Hobbs}, {Klioner}, {Lammers}, {McMillan}, {Ramos-Lerate},
  {Steidelm{\"u}ller}, {Stephenson}, \& {van Leeuwen}}]{LindegrenEDR3Astr2020}
{Lindegren}, L., {Bastian}, U., {Biermann}, M., {et~al.} 2021{\natexlab{a}},
  A\&A, 649, A4

\bibitem[{{Lindegren} {et~al.}(1992){Lindegren}, {H{\o}g}, {van Leeuwen},
  {Murray}, {Evans}, {Penston}, {Perryman}, {Petersen}, {Ramamani}, {Snijders},
  {S\"{o}derhjelm}, {Andreasen}, {Cruise}, {Elton}, {Lund}, \&
  {Poder}}]{lindegren+1992}
{Lindegren}, L., {H{\o}g}, E., {van Leeuwen}, F., {et~al.} 1992, A\&A, 258, 18

\bibitem[{{Lindegren} {et~al.}(2021{\natexlab{b}}){Lindegren}, {Klioner},
  {Hern{\'a}ndez}, {Bombrun}, {Ramos-Lerate}, {Steidelm{\"u}ller}, {Bastian},
  {Biermann}, {de Torres}, {Gerlach}, {Geyer}, {Hilger}, {Hobbs}, {Lammers},
  {McMillan}, {Stephenson}, {Casta{\~n}eda}, {Davidson}, {Fabricius},
  {Gracia-Abril}, {Portell}, {Rowell}, {Teyssier}, {Torra}, {Bartolom{\'e}},
  {Clotet}, {Garralda}, {Gonz{\'a}lez-Vidal}, {Torra}, {Abbas}, {Altmann},
  {Anglada Varela}, {Balaguer-N{\'u}{\~n}ez}, {Balog}, {Barache}, {Becciani},
  {Bernet}, {Bertone}, {Bianchi}, {Bouquillon}, {Brown}, {Bucciarelli},
  {Busonero}, {Butkevich}, {Buzzi}, {Cancelliere}, {Carlucci}, {Charlot},
  {Cioni}, {Crosta}, {Crowley}, {del Peloso}, {del Pozo}, {Drimmel}, {Esquej},
  {Fienga}, {Fraile}, {Gai}, {Garcia-Reinaldos}, {Guerra}, {Hambly}, {Hauser},
  {Jan{\ss}en}, {Jordan}, {Kostrzewa-Rutkowska}, {Lattanzi}, {Liao}, {Licata},
  {Lister}, {L{\"o}ffler}, {Marchant}, {Masip}, {Mignard}, {Mints}, {Molina},
  {Mora}, {Morbidelli}, {Murphy}, {Pagani}, {Panuzzo}, {Pe{\~n}alosa Esteller},
  {Poggio}, {Re Fiorentin}, {Riva}, {Sagrist{\`a} Sell{\'e}s}, {Sanchez
  Gimenez}, {Sarasso}, {Sciacca}, {Siddiqui}, {Smart}, {Souami}, {Spagna},
  {Steele}, {Taris}, {Utrilla}, {van Reeven}, \&
  {Vecchiato}}]{2021A&A...649A...2L}
{Lindegren}, L., {Klioner}, S.~A., {Hern{\'a}ndez}, J., {et~al.}
  2021{\natexlab{b}}, \aap, 649, A2

\bibitem[{{Lindegren} {et~al.}(2012){Lindegren}, {Lammers}, {Hobbs},
  {O'Mullane}, {Bastian}, \& {Hern{\'a}ndez}}]{lindegrenAGIS2012}
{Lindegren}, L., {Lammers}, U., {Hobbs}, D., {et~al.} 2012, A\&A, 538, A78

\bibitem[{Perryman \& Schulze-Hartung(2011)}]{Perryman+Schulze-Hartung2011}
Perryman, M. A.~C. \& Schulze-Hartung, T. 2011, A\&A, 595, A65

\bibitem[{Poisson \& Will(2014)}]{Poisson+Will2014}
Poisson, E. \& Will, C.~M. 2014, Gravity: Newtonian, post-Newtonian and
  relativistic (Cambridge: Cambridge Univ. Press)

\bibitem[{{Soffel}(1989)}]{1989racm.book.....S}
{Soffel}, M.~H. 1989, {Relativity in Astrometry, Celestial Mechanics and
  Geodesy} (Springer-Verlag)

\bibitem[{{Taylor}(2005)}]{Taylor2005}
{Taylor}, M.~B. 2005, in Astronomical Society of the Pacific Conference Series,
  Vol. 347, Astronomical Data Analysis Software and Systems XIV, ed.
  P.~{Shopbell}, M.~{Britton}, \& R.~{Ebert}, 29

\bibitem[{Vecchiato {et~al.}(2018)Vecchiato, {Bucciarelli, B.}, {Lattanzi, M.
  G.}, {Becciani, U.}, {Bianchi, L.}, {Abbas, U.}, {Sciacca, E.}, {Messineo,
  R.}, \& {De March, R.}}]{Vecchiato+2018}
Vecchiato, A., {Bucciarelli, B.}, {Lattanzi, M. G.}, {et~al.} 2018, A\&A, 620,
  A40

\bibitem[{{Will}(2014)}]{2014LRR....17....4W}
{Will}, C.~M. 2014, Living Reviews in Relativity, 17, 4

\end{thebibliography}

\end{document}